\documentstyle[prb,aps,epsf,rotate,color]{revtex}

\begin{document}
\twocolumn[\hsize\textwidth\columnwidth\hsize\csname @twocolumnfalse\endcsname

\title{
\vspace*{-10mm}
\hspace*{-9cm}{\small To appear in Phys. Rev. {\bf B58}, July 1, 1998.}\\
\vspace*{10mm}
\rule[9mm]{0in}{9mm} Suppression of Superconductivity
                     in UPt$_3$ Single Crystals}
\author{J.B. Kycia$^1$, J.I. Hong$^2$, M.J. Graf$^{1,\dag}$, 
J.A. Sauls$^1$, D.N. Seidman$^2$ and W.P. Halperin$^1$}

\address{$^1$Department of Physics \& Astronomy\\
	 $^2$Department of Materials Science and Engineering\\
         Northwestern University, Evanston, Illinois-60208}
\date{\today}
\maketitle

\begin{abstract}
High quality single crystals of UPt$_3$ have been prepared by vertical
float-zone refining in ultra-high vacuum.  We find 
that the superconducting transition temperature can be varied
systematically by annealing, revealing that the transition temperature
intrinsic to UPt$_3$ is $563\pm 5$~mK.
The suppression of the superconducting transition from defects is
consistent with a modified Abrikosov-Gor'kov formula that includes
anisotropic pairing, Fermi surface anisotropy, and anisotropic 
scattering by defects.
\newline 
\end{abstract}
] 

Since discovery of unconventional pairing in superfluid
$^3$He there has been considerable 
interest in the existence of similar states
in superconductors. Particular attention has been directed toward the
cuprates,\cite{sca95} some organic superconductors\cite{lee97}, and 
heavy fermion compounds.\cite{hef96}  These strongly correlated 
fermion superconductors have common
features, one of which is their sensitivity to elastic scattering from
defects and impurities that depends on the pairing
symmetry. In this work we focus attention on one of the most
intensively studied heavy fermion superconductors, UPt$_3$,
where we have developed a technique to systematically 
control the degree of elastic scattering.

In early transport experiments there were indications of
unconventional superconducting behavior from power law
temperature dependences of the attenuation of sound.\cite{bis84,shi86a}
Acoustic experiments\cite{mul87,qia87}
also showed the first evidence for multiple superconducting phases in a 
magnetic field. The observation\cite{fis89}
of a second superconducting transition 
in zero field heat capacity of UPt$_3$ led to a number of
studies of these phases and their
boundaries.\cite{has89,bru90,ade90}  UPt$_3$ has a
rich phase diagram in the field-temperature plane with at least three
distinct phases at fields below the upper critical field. The
existence of these phases is strong evidence
for a multicomponent order parameter.

Theoretical models based on several different symmetry classes
have been proposed for the 
multi-component order parameter.\cite{sau94,hef96}
Experimental and theoretical work on both the phase diagram and
transport properties has narrowed the number of viable models,
but currently the precise symmetry
class for the pairing state of UPt$_3$ is not settled.
Because the signatures of the symmetry of the pairing state
are particularly sensitive to material quality,
one of the key experimental challenges is the preparation and
characterization of high-purity single
crystals with which to test various predictions.
Theoretical work on the thermodynamic and transport properties
of unconventional superconductors shows that a powerful way
of testing different candidate pairing states is to study the effects
of impurity scattering on physical properties of the superconducting
state.\cite{muz93} For example, it has been shown that the 
limiting values for the components of the thermal conductivity
tensor, e.g., $\lim_{T\rightarrow 0}\kappa_{a,c}/T$,
are sensitive to impurities in very different ways depending
on the orbital symmetry of the pairing state.\cite{gra96}

It is well-known that for s-wave pairing in isotropic
superconductors, scattering by non-magnetic impurities
has no effect on $T_c$.\cite{abr59a,and59}
However, the transition temperature for
unconventional superconductors
is particularly sensitive to
scattering by non-magnetic impurities or defects.\cite{gor87}
The first observation of this effect was by Stewart et al.\cite{ste84}
In recent work on polycrystalline samples, Dalichaouch et al.\cite{dal95}
substituted a variety of elements for uranium in UPt$_3$ and 
demonstrated the suppression of $T_c$ with residual resistivity.
These authors argue that the rare earth impurities produce
magnetic and non-magnetic scattering, and that their results are consistent
with the behavior expected for unconventional superconductors.
In earlier impurity studies, Vorenkamp et al.\cite{vor93}
substituted Th and Y for U, and obtained 
qualitatively similar results.
Ideally the study of the suppression of $T_c$
should be performed on single crystals.
Grain boundaries and the anisotropy of UPt$_3$, and possibly
the anisotropy of scattering processes, complicate the interpretation
of the experimental results in polycrystalline samples.

We report the first study of the effects of impurity scattering on the
suppression of superconductivity in single crystals of UPt$_3$.
As a test for unconventional pairing it is important to measure the
suppression of superconductivity attributable to non-magnetic scattering.
We have found a means to systematically
control defect concentrations, and we have
determined their influence on elastic scattering and
suppression of superconductivity in single crystals.

The crystals we prepared
have sufficient purity and perfection that we can reliably
extrapolate to the clean limit and obtain, for the first
time, the
intrinsic superconducting transition temperature of UPt$_3$,
$563\pm 5$~mK.
From our 
measurements (see below) we
conclude that the defects are not chemical impurities.
The observed suppression of the transition
temperature with increased elastic scattering from these defects is
further evidence for unconventional pairing in this compound.

The crystals were prepared in a vertical float-zone refining system
that operates with electron-beam heating in ultra-high vacuum 
($10^{-10}\,\mbox{torr}$).\cite{kyc95}
Annealing was performed at different
temperatures in a furnace, also operating in ultra-high vacuum, where
the sample was placed on a UPt$_3$ holder and heated by an electron gun for
six days followed by slow cooling over a four day period.  Characterization
by X-ray rocking curves  and chemical analysis based on mass spectrometry
were performed.\cite{kyc95}
The total impurity concentration was determined to be $10$ parts per
million by weight ($23$ ppm atomic concentration) and the concentrations of
conventional magnetic impurities (Fe, Cr, Co, Mn, Ni) are less than 
$0.03$ ppm by weight.

The resistivity, susceptibility and specific heat were measured
to temperatures well below the superconducting transition. For all samples
studied the specific heat exhibited a well 
defined double peak.\cite{kyc95}
The X-ray rocking curve of the $[002]$ reflection,
for a sample annealed at $1250^{\rm o}\mbox{C}$ had
a FWHM of $63 \pm 5$ arcseconds compared to the theoretical minimum 
FWHM calculated to be 60 arcseconds according to the
Darwin-Prins model.\cite{war69}  This is evidence for very high crystal
quality. The samples which we report on
here were obtained from three zone refining runs, which 
produced large single crystals ($\sim 20\,\mbox{g}$).
Rectangular, needle-shaped specimens, typically 
$\approx 5 \times 0.5 \times 0.5\,\mbox{mm}^3$,
were cut by electron discharge machining. The transition temperature
was determined by an a.c.\ four-probe technique with excitation
$< 30\,\mu\mbox{A}$, to avoid self-heating.\cite{kyc95}

In Fig.~1 we show the behavior of the residual electrical
resistivity (RRR), extrapolated to zero temperature, 
for different annealing temperatures.
Note that each point represents
a separate sample.
The RRR is expressed as a ratio relative to the
room temperature resistance. The resistivity of UPt$_3$ is anisotropic and so
we have used the values $\rho_{\mbox{\tiny 0}a}/\rho_{\mbox{\tiny 0}c}=3.5$
at $T=0$, from Fig.~4, and
$\rho_a/\rho_c=1.80$ at room temperature,\cite{dev84}
to express the measurements for $a$-axis orientations in terms 
of the resistivity for the $c$-axis
orientation, defined here as RRR$_c$. The general trend of
increasing RRR with decreasing annealing temperature is evident.  The
difficulty in establishing and measuring the annealing conditions, including
the cooling schedule, likely gives rise to the spread in the data in Fig.~1
for low annealing temperatures.
Nonetheless, a clear trend is apparent and the best quality
crystals have RRR$_c$ of order $1500$, in
comparison with bulk crystals reported
in the literature, RRR$_c < 750$.\cite{kim95,lus94,sud97}
From previous work we estimate
that contributions to the c-axis
resistivity from chemical impurities are less than
$14\,\mu\Omega\, \mbox{cm/atomic}\%$.\cite{dal95}
We use our impurity concentration values to establish (quite
conservatively) that they would limit the RRR$_c$ to be above
$4000$, giving a negligible contribution
to the measured residual resistivities of our crystals.
Using transmission electron microscopy we find planar defects lying in the 
basal planes with the defect density being higher for the higher resistivity 
samples.\cite{hon97}  The measured density of 
planar defects is consistent with the measured residual resistivities.
Thus, we conclude that the residual resistivity has a significant
contribution from these structural defects.

\begin{figure}
\centerline{
\epsfysize=0.45\textwidth
\rotate[r]{\epsfbox{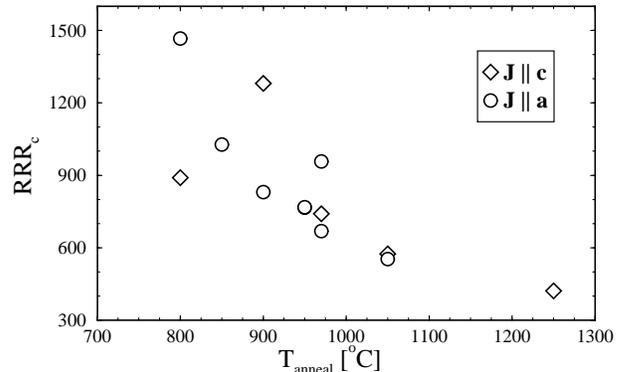}}
}
\begin{minipage}{0.45\textwidth}
\caption[]{RRR$_c$ from all the crystals.
The measured residual resistivity
ratio for current, $J$, in the $c$-axis direction, $J || c$; 
along with the equivalent values of RRR$_c$
inferred from the $J || a$ data.}
\end{minipage}
\end{figure}

In Fig.~2 we show the resistive transitions of
two crystals. The inset shows one of the narrowest resistive
transitions measured to date, $\Delta T_c=1.3\,\mbox{mK}$, determined as the
interval between $10\%$ to $90\%$ of the jump in $\rho_0$.

\begin{figure}
\centerline{
\epsfysize=0.45\textwidth
\rotate[r]{\epsfbox{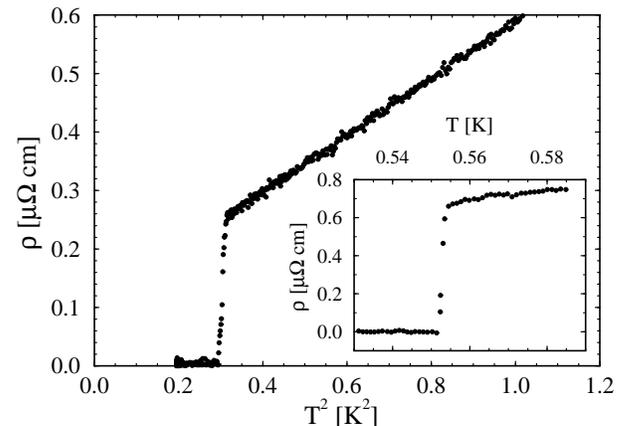}}
}
\begin{minipage}{0.45\textwidth}
\caption[]{Resistivity for a sample annealed at 
$900^{\rm o}\mbox{C}$ and measured with $J || c$. 
Inset: Superconducting transition for a sample annealed at 
$800^{\rm o}\mbox{C}$ with $J || a$
having a transition width of $1.3\,\mbox{mK}$.}
\end{minipage}
\end{figure}

The anisotropic resistivity has two components at low temperatures, Fig.~2,
\begin{equation}
\rho_i(T) = \rho_{\mbox{\tiny 0}i}+A_i\,T^2\,.
\end{equation}
The index, $i=(a,c)$, identifies the
direction of the current relative to the hexagonal unit
cell. The temperature independent term, $\rho_{\mbox{\tiny 0}i}$,
results from elastic
scattering of quasiparticles from defects and impurities,
and the $T^2$ term is due
to inelastic scattering, which we assume is
a result of quasiparticle-quasiparticle
collisions. The coefficient $A_i$ is then inversely
proportional to the average squared Fermi velocity projected along the
direction $i$. As a consequence of the large effective 
masses, the $A_i$ coefficients in heavy fermions are
much larger than in conventional metals. 
The results of our measurements of the anisotropy of the elastic scattering
coefficient are presented in Fig.~3.
Note that the inelastic coefficients are independent of 
the residual resistivity, as expected for quasiparticle-quasiparticle
scattering with low concentrations of defects and impurities.

\begin{figure}
\centerline{
\epsfysize=0.45\textwidth
\rotate[r]{\epsfbox{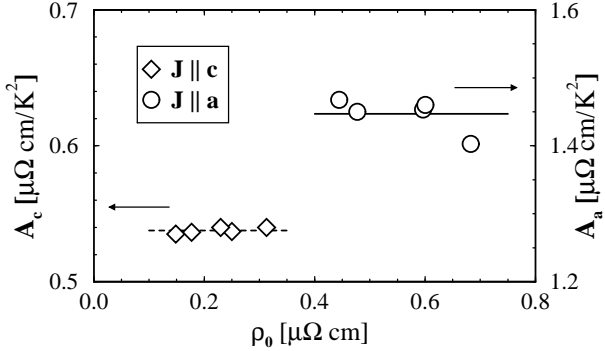}}
}
\begin{minipage}{0.45\textwidth}
\caption[]{Inelastic scattering coefficients for both orientations 
as a function of $\rho_{\mbox{\tiny 0}}$.
}
\end{minipage}
\end{figure}

If the inelastic scattering probability is isotropic,
the anisotropy in the angular average of the Fermi
velocity can be determined from the
$A_i$ coefficients,
\begin{equation}
\frac{A_a}{A_c}=\frac{\langle v_{fc}^2(\bbox{p}_f) \rangle}{
  \langle v_{fa}^2(\bbox{p}_f) \rangle}=2.69 \pm 0.03
\,.
\end{equation}

The elastic scattering rate can be inferred from 
the residual resistivities, $\rho_{\mbox{\tiny 0}i}$,
and the Drude result,
\begin{equation}\label{Drude}
\rho_{\mbox{\tiny 0}i}^{-1}=\frac{3}{\pi^2}\,
\left(\frac{e}{k_{\tiny B}}\right)^2\gamma_{\tiny S}
\left< v_{fi}^2(\bbox{p}_f)\,\tau(\bbox{p}_f) \right>
\,,
\end{equation}
where $\tau(\bbox{p}_f)$ is the transport
time for quasiparticles scattering into
the direction $\bbox{p}_f$ on the Fermi surface, and 
$\gamma_{\tiny S}=\frac{2}{3}\pi^2 k_B^2 N_f$
is the Sommerfeld coefficient for the linear term in the electronic 
specific heat. Thus, the anisotropy in the residual resistivity,
$\rho_{\mbox{\tiny 0}a}/\rho_{\mbox{\tiny 0}c}$, is determined 
by the anisotropy in the Fermi velocities and the elastic scattering time.
We define effective scattering times for $a$-axis and $c$-axis transport
by
$\tau_i= \big\langle v_{fi}^2(\bbox{p}_f)\,\tau(\bbox{p}_f) \big\rangle /
\big\langle v_{fi}^2(\bbox{p}_f) \big\rangle$.
For isotropic scattering times the ratio of residual
resistivities for $a$-axis and $c$-axis measurements is inversely
proportional to the ratio of Fermi-surface averages of
the $ab$-plane and $c$-axis Fermi velocities.
However, the Fermi velocity anisotropies above cannot account for the 
anisotropy of $\rho_{\mbox{\tiny 0}i}$.  From our measurements of the 
suppression of $T_c$ as a function of residual resistivity we infer that 
the elastic scattering rate is anisotropic in UPt$_3$, i.e.,
\begin{equation}\label{tau-anisotropy}
\frac{\tau_c}{\tau_a}=
\frac{\rho_{\mbox{\tiny 0}a}\,A_c}{\rho_{\mbox{\tiny 0}c}\,A_a}
= 1.3\pm 0.1
\,.
\end{equation}
The ratio $\rho_{\mbox{\tiny 0}a}/\rho_{\mbox{\tiny 0}c}=3.5 \pm 0.3$ 
used in Eq.~(\ref{tau-anisotropy}) is obtained
from the slopes of the $T_c$ suppression for $a$-axis and $c$-axis 
currents in Fig.~4. Extrapolation to zero resistivity gives
$T_{c0}=563\pm 5\,\mbox{mK}$, where the accuracy is
determined by absolute thermometry.

\begin{figure}
\centerline{
\epsfysize=0.45\textwidth
\rotate[r]{\epsfbox{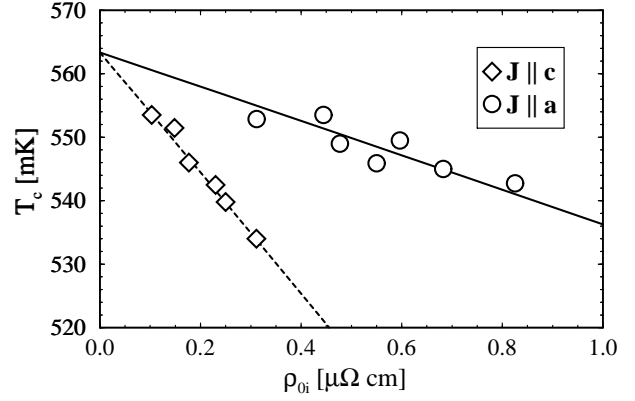}}
}
\begin{minipage}{0.45\textwidth}
\caption[]{
$T_c$ suppression in UPt$_3$ vs. the residual resistivity
for both orientations. 
The solid and dashed lines are theoretical fits
for an $E_{1g}$ or $E_{2u}$ order parameter (see below).
}
\end{minipage}
\label{figure}
\end{figure}

The basis for this analysis
follows from a generalization of the Abrikosov-Gor'kov formula for
the pair-breaking suppression of $T_c$ to include anisotropic scattering
in unconventional superconductors.
The suppression of $T_c$ by a homogeneous distribution of
isotropic, non-magnetic impurities
has been calculated for unconventional
superconductors,\cite{gor87}
and is given by a formula similar to the
Abrikosov-Gor'kov formula 
for pair-breaking in conventional superconductors
by magnetic impurities, i.e., 
$\ln(T_{c0}/T_{c})=\Psi(\mbox{$1\over 2$}+
 (\hbar/2\pi\tau k_{\tiny B} T_c))
-\Psi(\mbox{$1\over 2$})$,
where $1/\tau$ is the isotropic impurity scattering rate, 
$T_c$ is the superconducting transition temperature,
$T_{c0}$ is the transition temperature of the perfect,
clean material, and $\Psi$ is the digamma function.
This equation, combined with the Drude formula, relates the suppression of
$T_c$ to the residual resistivity, $\rho_{\mbox{\tiny 0}}$.
We extend this result to include anisotropic scattering
in superconductors with an unconventional order parameter 
and an anisotropic Fermi surface.
The simplest model that can account for the
observed uniaxial anisotropy of the resistivity
is a two parameter model (s-wave and p-wave)
for the scattering rate
of a quasiparticle with Fermi momentum $\bbox{p}_f$
scattering off anisotropic impurities or defects,
\begin{equation}\label{sp-model}
1/\tau(\bbox{p}_f)=1/\tau_{\mbox{\tiny 0}}
-3\hat{p}_{fz}^2/\tau_{\mbox{\tiny 1}}
\,,
\end{equation}
where $\hat{p}_f$ is the direction of the Fermi momentum and
$1/\tau_{\mbox{\tiny 0}}\ge 3/\tau_{\mbox{\tiny 1}}$ guarantees
$1/\tau(\bbox{p}_f)\ge 0$.
This model is applicable to anisotropic scattering by
a nearly homogeneous distribution of structural defects.\cite{thu98}
The suppression of T$_c$ also depends on the anisotropy
of the pairing state. We restrict the analysis to the pairing
states that can explain both the H-T phase diagram\cite{hef96,sau94}
and the anisotropic thermal conductivity.\cite{gra96}
This restriction allows only the 
order parameters belonging to the $E_{1g}$ 
or $E_{2u}$ pairing symmetries, which are the leading candidates for
the superconducting phases of UPt$_3$.\cite{hef96,sau94}
The ground state basis functions for
these two models are $\eta^{\pm}_{E_{2u}}(\bbox{p}_f) = \hat p_{fz}
(\hat p_{fx}\pm i \hat{p}_{fy})^2$ and
$\eta^{\pm}_{E_{1g}}(\bbox{p}_f) = \hat{p}_{fz}
(\hat{p}_{fx}\pm i \hat{p}_{fy})$.
The calculation of the suppression of T$_c$,\cite{gor87}
including anisotropic scattering and anisotropic pairing, gives 
\begin{eqnarray}\label{formula}
\ln \frac{T_c}{T_{c0}} = 
	\Psi \left( \mbox{$1\over 2$} \right) -
	{\cal Q}^{-1}
	\left \langle |\eta(\bbox{p}_f)|^2
	 \Psi \left( \mbox{$1\over 2$}
	   +\mbox{$1\over 2$} \alpha(\bbox{p}_f) \right)
	\right \rangle	\,,
\end{eqnarray}
where $\alpha(\bbox{p}_f) = \hbar/(2\pi k_{\tiny B} T_c \tau(\bbox{p}_f))$ 
and ${\cal Q} = \langle |\eta(\bbox{p}_f)|^2 \rangle$.
For weak scattering Eq.~(\ref{formula}) reduces to
\begin{equation}\label{expansion}
\frac{T_c}{T_{c0}} \simeq 1 - \frac{\pi^2}{4}
	{\cal Q}^{-1}
	\left\langle 
	  |\eta(\bbox{p}_f)|^2 \alpha(\bbox{p}_f) 
	\right\rangle
\,.
\end{equation}
Evaluating the Fermi surface averages
for the s-p model gives\cite{footnote}
\begin{equation}
\frac{T_c}{T_{c0}} \simeq 1 - \frac{\pi^2}{4} q\, \alpha_{0}
\,,
\quad \alpha_0 = \frac{\hbar}{2\pi k_{\tiny B} T_{c0} \tau_{\mbox{\tiny 0}}}
\,,
\end{equation}
where $q=1-1/3r$ for the $E_{2u}$ model and $q=1-3/7r$
for the $E_{1g}$ model.  The anisotropy ratio is defined 
by $r=\tau_{\mbox{\tiny 1}}/3\tau_{\mbox{\tiny 0}}$.

Similarly, from the s-p model for the scattering rate and
the definition of the effective scattering times, we can relate these
to the s-wave and p-wave scattering times,
\begin{eqnarray}
&&\tau_{a}  = \tau_{\mbox{\tiny 0}}\,
  \frac{r}{2}
  \left(
     1 - \left( \sqrt{r}-{1}/{\sqrt{r}} \right)
     \,\tanh^{-1} \left( {1}/{\sqrt{r}}	\right)
  \right)	\,,
\\ &&
\tau_{c} = \tau_{\mbox{\tiny 0}}\,
  r
  \left(
     \sqrt{r} \,\tanh^{-1} \left( {1}/{\sqrt{r}} \right) - 1	
  \right)	\,.
\end{eqnarray}
These results for $T_c/T_{c0}$ and $\tau_{a,c}$ can be combined to express 
$T_c/T_{c0}$ in terms of either $\rho_{\mbox{\tiny 0}a}$ or 
$\rho_{\mbox{\tiny 0}c}$. 
The slopes, 
$dT_c/d\rho_{\mbox{\tiny 0}a} = -26.6 \pm 2 \, \mbox{mK/$\mu\Omega\,$cm}$, 
$dT_c/d\rho_{\mbox{\tiny 0}c} = -93.0 \pm 1 \, \mbox{mK/$\mu\Omega\,$cm}$, 
from Fig.~4 can be used to determine $\tau_a$ and $\tau_c$,
and thus the s- and p-wave scattering times.  From the measured anisotropy 
ratios we obtain $\tau_c/\tau_a=1.3\pm 0.1$ [Eq.~(\ref{tau-anisotropy})].
The ratio of the s-wave and p-wave scattering rates 
is then $r\simeq 2$, which corresponds to effective scattering times 
$\tau_a\simeq 1.13\tau_{\mbox{\tiny 0}}$ and 
$\tau_c\simeq 1.48\tau_{\mbox{\tiny 0}}$.  
Combined with the Sommerfeld coefficient,
$\gamma_{\tiny S}=430\, {\rm mJ\, mol^{-1}\, K^{-2}}$,\cite{fis89}
and Eq.~(\ref{Drude}), we obtain 
$\langle v_{f}^2(\bbox{p}_f) \rangle^{1/2} 
= (2 \langle v_{fa}^2\rangle + \langle v_{fc}^2\rangle)^{1/2}
\simeq 3.3\,\mbox{km/s}$.
This average is consistent with averaged velocities for extremal orbits 
obtained from de Haas-van Alphen measurements,
$3.5-5.5\,\mbox{km/s}$.\cite{tai87}

In conclusion we have grown high quality crystals of UPt$_3$ and have
found a means for controlling the elastic scattering by ultra-high vacuum
annealing. Using transmission electron microscopy
we have identified the defects principally responsible for the scattering
as planar defects. We show independently that they
are not chemical impurities. The measured suppression
of $T_c$ with increasing residual resistivity is in good
agreement with a simple generalization of the Abrikosov-Gor'kov theory,
which includes anisotropic scattering, unconventional pairing,
and Fermi surface anisotropy. We infer from the data and this model
that the effective scattering rate is approximately
$30\%$ weaker for c-axis transport compared with $ab$-plane transport.
Finally, we have shown that our crystals approach the perfect,
clean limit under appropriate annealing conditions. 
By extrapolating to zero elastic
scattering we deduce the intrinsic transition temperature
of UPt$_3$ to be $563\pm 5\,\mbox{mK}$.

We thank M. Bedzyk, B. Davis and M. Meisel for their contributions
to this work and to the heavy fermion research project at Northwestern.
The research was supported in part by the NSF (DMR-9705473) Focused
Research Grant Program and the NSF (DMR-9120000) Science and Technology
Center for Superconductivity and by the NEDO Foundation.  Use was made
of the Central Facilities of the Northwestern University Materials
Research Center supported by NSF (DMR-9120521) MRSEC Program.


\begin{thebibliography}{10}
\bibitem{footnote}
For calculating the Fermi surface averages we mapped an ellipsoidal
Fermi surface onto a sphere.
\end{thebibliography}
\end{document}